\documentclass[review]{elsarticle}

\usepackage{setspace}
\usepackage{geometry}
\usepackage{pdflscape}
\usepackage{amsmath, amsthm, amssymb, mathrsfs}
\newtheorem{hyp}{Hypothesis}
\usepackage{bbm}

\makeatletter
\def\ps@pprintTitle{%
   \let\@oddhead\@empty
   \let\@evenhead\@empty
   \let\@oddfoot\@empty
   \let\@evenfoot\@oddfoot
}
\makeatother

\biboptions{authoryear, sort}

\def\sym#1{\ifmmode^{#1}\else\(^{#1}\)\fi}

\begin{document}

\begin{frontmatter}

\title{Technological impact of biomedical research: the role of basicness and novelty}

\author{Qing Ke\corref{corrauthor}}
\address{Northeastern University, Boston, MA 02115, USA}
\cortext[corrauthor]{Corresponding author}
\ead{q.ke@northeastern.edu}

\begin{abstract}
An ongoing interest in innovation studies is to understand how knowledge generated from scientific research can be used in the development of technologies. While previous inquiries have devoted to studying the scientific capacity of technologies and institutional factors facilitating technology transfer, little is known about the intrinsic characteristics of scientific publications that gain direct technological impact. Here we focus on two features, namely basicness and novelty. Using a corpus of 3.8 million papers published between 1980 and 1999, we find that basic science papers and novel papers are substantially more likely to achieve direct technological impact. Further analysis that limits to papers with technological impact reveals that basic science and novel science have more patent citations, experience shorter time lag, and have impact in broader technological fields.
\end{abstract}

\begin{keyword}
patent-to-paper citation \sep non-patent reference \sep technological impact \sep basic research \sep novelty
\end{keyword}

\end{frontmatter}

\section{Introduction}

It has been long argued and empirically demonstrated that scientific research spurs technological innovation, promotes economic growth, and improves human health. One line of inquiry has emphasized establishing the intimate interplay between science and technology \citep{Jaffe-real-1989, Cockburn-public-1996, Godin-research-1996, Narin-linkage-1997, McMillan-biotech-2000, Meyer-patent-2001, Breschi-tracing-2010, Han-testing-2018}. An early study by \citet{Jaffe-real-1989}, for example, found a significant spillover effect of university research on corporate patents. Another stream of works from a range of fields has provided abundant evidence on the effects of publicly funded biomedical research on the improvement of health outcomes \citep{Comroe-scientific-1976, Cockburn-public-1996, Cutler-return-2003, Manton-nih-2009, Morlacchi-how-2011, Sampat-what-2011, Stevens-role-2011, Blume-Kohout-does-2012}. Focusing on cardiovascular disease as a case study, \citet{Cutler-return-2003} estimated a 4-to-1 return for investments in new treatments and 30-to-1 return for investments in health knowledge. These lines of researches together substantiate the effects of scientific research, providing basis for the advocation of public support for science. 

Accompanying ample studies of the effects of scientific research is the understanding of the mechanisms and channels through which such effects are realized. Among the diverse channels, a widely recognized one is new technologies, and revealing how knowledge produced from scientific research can be used in the development of technologies is an important topic in the innovation study literature. Prior literature on this topic has suggested numerous institutional factors that may facilitate the science-to-technology translation process, from the perspectives of both the university side \citep{Etzkowitz-dynamics-2000, Gregorio-why-2003, Debackere-role-2005} and the private sector side \citep{Gambardella-competitive-1992, Cockburn-public-1996, Laursen-open-2006}.

A less explored perspective, which is the focus of this work, is science itself. Here we ask: Which characteristics of scientific publications make them more likely to be cited in front-pages of patented technologies (also known as ``prior art'')? We are primarily interested in two features, namely basic science and novel science.

Our interest in basic science comes from its role as ``the pacemaker of technological progress.'' In a 1945 report to the President of the United States of America titled \emph{Science: The Endless Frontier} \citep{Bush-science-1945}, Vannevar Bush made a distinction between basic and applied research and claimed an essential role of basic science in technological innovation and economic growth. He argued that, in order to promote growth, it is government's concern to support and sustain basic research. In our focused area of biomedical research, numerous studies have established that basic science lays the foundations for therapeutic applications that improve human health \citep{Comroe-scientific-1976, Cutler-return-2003, Stevens-role-2011}. In this work, we add to another dimension of the impact of basic science by empirically investigating whether it is more likely to attain direct technological impact, as defined as being referenced as ``prior art'' by patented technologies.

Novelty is the other feature studied in this work because of its particularity of high risk and high return. Novel science, when combined with high conventionality, has been shown to exhibit particular advantage to be scientific breakthroughs \citep{Uzzi-atypical-2013}, but at the same time encounters resistance of acceptance from existing paradigms and thus experiences delay recognition from the scientific community. In biomedical research, it has been reported that researchers tend to adopt conservative approaches \citep{Rzhetsky-choosing-2015}, and funding agencies tend to favor safe project proposals rather than novel and risk ones that explore uncharted areas, although funding policy is one of the key drivers of scientific creativity \citep{Azoulay-incentives-2011}. Here we ask: Is novel science more likely to generate technological impact? Does novel science also face delayed recognition in the technology sphere like in the scientific community?

To answer these questions, we draw on two large-scale corpora: (1) 3.8 million papers published between 1980 and 1999 and indexed in the MEDLINE database, and (2) all utility patents granted at the U.S. Patent and Trademark Office (USPTO) until 2012. We show that basic science papers are more likely to achieve technological impact than comparable clinical research papers in the same field and the same year, and the effect size is large. This positive linkage to technological impact also holds for novel science. We further find that both positive linkages are robust even after controlling for a range of confounders like the quality of papers and journal Impact Factor, and consistent during the studied 20-year period. Additional analysis on the subset of papers that are cited by patents reveals that basic science and novel science have more patent citations and a broader scope of technological impact. Moreover, it takes less time for both of them to have impact in the technology space.

\section{Literature review}
\label{sec:lit}

\subsection{Technological impact of scientific research}

Tracing and understanding how scientific research contribute to the development of technologies has been an important question in the literature of innovation studies. Following the seminal work of \citet{Narin-linkage-1997}, empirical measurement of science-technology interaction has largely relied on non-patent references (NPR) that are listed in the front-pages of patents and refer to scientific publications. Hereafter, we call those references scientific NPRs, or SNPRs. They are, however, not the only type of data used for studying the interaction. Some existing works, for example, have used references in the full-text of patents \citep{Bryan-text-2019}, or have employed natural language processing techniques to understand text of scientific papers and patents \citep{Xu-novel-2019}. In this work, we follow the literature that uses front-page SNPRs.

What do SNPRs represent? How can one interpret them? These questions are still in discussion. One of the most popular interpretations is that SNPRs signal knowledge flow from science to technology. \citet{Narin-linkage-1997} observed a rapid increase of the volume of citations from U.S. patents to public science, defined as research conducted at universities and public research organizations, and argued that there was a growing reliance of technologies on public science. In the follow-up work, \citet{McMillan-biotech-2000} examined SNPRs in patents from the biotechnology industry and found that its reliance on public science is more evident than other industries.

A limited number of studies instead set out to examine the validity of representing SNPRs as knowledge flows. \citet{Meyer-does-2000} studied front-page SNPRs of several patents about nanoscale technologies and suggested that the interpretation of a direct knowledge link from cited papers to citing patents is hardly guaranteed and scientific papers play an indirect role by providing background information. \citet{Callaert-source-2014} conducted interviews with inventors at Belgian firms and institutes and reported two observations: (1) 30\% of patents with scientific knowledge as a source of inspiration do not have any SNPRs; (2) half of SNPRs served as unimportant or background information. They suggested that the knowledge flow interpretation is not guaranteed and SNPRs may also signal relatedness (e.g., providing background information) between citing patent and cited paper. \citet{Roach-lens-2013} termed the first observation as ``errors of omission'' and examined another type of distortion called ``errors of commission''---citations that contributed to the invention were not reported. They used the Carnegie Mellon Survey of the Nature and Determinants of Industrial R\&D where lab managers reported how knowledge originated from public research organizations was used in R\&D. Their analysis suggested that citations to scientific publications more align with reports from managers than citations to patents; that is, compared with representing backward patent citations as knowledge flow, SNPRs better represent this notion.

Regardless of the extent of the validity of the knowledge flow interpretation, SNPRs provide a valuable lens through which one can understand the linkage from science to technology. Many studies towards this goal take the \emph{patent}-centric perspective. \citet{Fleming-science-2004} proposed a mechanism through which science can contribute to technology is that science alters the combinatorial search process towards more useful combinations and corroborated the mechanism with empirical analysis showing that patents referencing science get more forward patent citations. \citet{Cassiman-search-2008} found that patents that have SNPRs have a broader scope of citations received from patents. In a large-scale study consisting of 22.9 million MEDLINE papers and 4.8 million U.S. patents, \citet{Ke-analysis-2019} observed an exponential growth of SNPRs, doubling every 2.9 years over a 23-year period (1976--1998).

The time dimension of technological impact is another topic of interest. Using a sample of U.S. patents from 1975 to 1989, \citet{Narin-status-1992} presented a \emph{retrospective} analysis of patent-to-paper citation linkage, finding that electronics and pharmaceuticals patents tend to cite more recent papers, with a median lag of 3--4 years. \citet{Ke-compare-2018} instead analyzed the time facet in a \emph{prospective} manner and found that, for the majority of biomedical papers, their citations from patents lag behind citations from scientific papers and the median lag is 6 years. 

Previous studies have also observed homophily in patent-to-paper citation linkage with respect to the field and country dimension. \citet{Narin-linkage-1997}, for example, observed an apparent subject-specific linkage to science; that is, drug and medicine patents mostly cite clinical medicine and biomedical research papers, and chemical patents cite chemistry papers. A similar observation is noted in \citet{Meyer-patent-2001}. In terms of the country aspect, \citet{Narin-linkage-1997} found that patents originated from one country tend to cite papers authored from the same country.

What remains relatively less studied is from the \emph{paper}-centric perspective. What are the characteristics of scientific papers that accrue technological impact? \citet{Ahmadpoor-dual-2017} presented an analysis of an integrated network formed by patent-to-patent, patent-to-paper, and paper-to-paper citations. They found that papers that are directly cited by patents are from fields like nanoscience and nanotechnology, computer science, etc. Mathematics papers, on the other hand, are distant to the technological space, meaning that several intermediate papers are needed in order to form a citation path from a patent to a mathematics paper. \citet{Ke-compare-2018} found that only 4\% of papers are directly cited by patents. Yet, the fraction varies by field; biotechnology, virology, biochemistry, molecular biology, allergy and immunology, and cell biology have more than 10\%, whereas medicine and general surgery have less than 2\%. Another studied characteristic is the novelty of papers. Using papers published in 2001, \citet{Veugelers-scientific-2019} found that within-filed novel papers are more likely to have patent citations.

While features examined in the above studies are of \emph{ex~ante}, other prior works have instead investigated \emph{ex~post} features, among which scientific impact is perhaps the one received the most attention. Studies have repeatedly established an overall positive association between scientific impact and technological impact; that is, the number of citations received by a paper from other scientific papers and from patents are positively correlated \citep{Ahmadpoor-dual-2017, Popp-from-2017, Ke-compare-2018}. However, \citet{Ke-compare-2018} further noted that within biomedicine, there is a low similarity between the set of top papers based on scientific impact and that based on the technological impact.

\subsection{Basic research and technological impact}

In this work, we are interested in the biomedicine area and aim to understand technological impact of biomedical research in terms of the paper-centric perspective. We study two kinds of paper characteristics---basicness and novelty---that are highly relevant to the current science policy discussions. We hypothesize that both of them may link to technological impact.

Our interest in basic science resides in its long-argued spillover effect in driving applied research and innovations. \citet{Bush-science-1945} argued that new knowledge generated from basic research is essential to technological and economic development and it is of government's concern to support and sustain basic research in order to promote such development. These arguments have later been absorbed into the widely discussed (and criticized) ``linear model'' of innovation \citep{Balconi-defence-2010}. Empirically, \citet{Narin-structure-1976} developed an indicator that categorized papers into four groups (basic research, clinical investigation, clinical mix, and clinical observation) based on journals where they were published. In their later work, \citet{Narin-linkage-1997} found that most papers that are cited by patents are from the basic research category. Using an indicator that quantifies the extent to which a biomedical paper is basic science or clinical medicine \citep{Ke-identify-2019}, \citet{Ke-analysis-2019} similarly found that patent-referenced papers are overwhelmingly from the basic science end of the basic-clinical spectrum. But for the subset of patents that are related to the development of FDA-approved drugs, a significant portion of their cited papers are located at the clinical research end of the spectrum \citep{Ke-analysis-2019}.

In light of these studies, our first hypothesis is:
\begin{hyp}
Basic science is more likely to have patent citations than comparable clinical research.
\end{hyp}

\subsection{Combinational novelty and impact}

We are interested in novel science because of its high-risk yet high-return particularity. Revolutionary progresses in science and technology often need to break away from existing accepted paradigms \citep{Kuhn-structure-1962} and tend to make novel combinations of previous components \citep{Lee-recombinant-2001}. Yet, the lack of fit to prevailing paradigms makes their recognition difficult and delayed. Indeed, novelty is an extensively studied mechanism of breakthrough recognition, and numerous studies have suggested that there is bias against novelty \citep{Boudreau-look-2016, Wang-bias-2017, Chai-breakthrough-2019}.

The notion of novelty from a combinational perspective is prevalent in the innovation literature. It has been studied for both scientific and technological innovations. Novel science has been operationalized as making combinations of existing knowledge components that have not been combined before. \citet{Uzzi-atypical-2013} viewed a scientific journal as one knowledge component and calculated a z-score for each journal pair by comparing the observed frequency in the bibliographies of papers published in a year with expected frequency by chance. A paper's conventionality and novelty are then defined as the median and 10th-percentile of the z-score distribution over its referenced journal pairs. Using the two indicators, they found that papers with the highest scientific impact are those that exhibit both high novelty and high conventionality. \citet{Foster-tradition-2015} looked at chemical-chemical combinations and found that papers examining new combinations of chemicals are more likely to have high impact than papers studying conservative chemical pairs. \citet{Boudreau-look-2016} treated a Medical Subject Heading (MeSH) term as a knowledge component and measured the novelty of a research proposal as the fraction of MeSH pairs that have not been used in the previous literature. They found that there is a systematic bias against novel proposals. \citet{Wang-bias-2017} quantified novelty by looking at pairs of referenced journals, and their analysis reinforced earlier findings about the bias against novel science.

Turning to the technology domain, \citet{Lee-recombinant-2001} found that patents that make new combinations have more variability, which can give rise to both success and failure in terms of forward patent citations. \citet{Sarah-double-2014} developed a novelty measure based on topic-modeling of text and found that new topic originated patents are featured with local search and both broader recombinations and novelty are linked to economic value of patents. \citet{Verhoeven-measuring-2016} measured technological novelty of patents with respective to both recombination and knowledge origin, as manifested by SNPRs, and found that technologically novel patents are more likely to receive high forward patent citations and at the same time exhibit higher dispersion in such citations.

Based on these works, our second hypothesis is:
\begin{hyp}
Novel science is more likely to have patent citations than comparable non-novel research.
\end{hyp}

\section{Data and Methods}
\label{sec:data}

\subsection{Sample selection}

To study the characteristics of scientific publications with technological impact, we link their information in the scientific domain with information in the technology space. The unit of our analysis is a scientific paper. The corpus of papers is those indexed in the MEDLINE database and published between 1980 and 1999. MEDLINE is among the most comprehensive bibliography datasets for the biomedical research literature. We obtain additional information, such as number of references and author affiliations, from the Web of Science (WoS) database. We only consider documents that are articles based on both the ``Publication Type'' tag in MEDLINE, as operationalized from https://icite.od.nih.gov/help, and document type specified in WoS. We further limit our analysis to papers whose WoS Subject Category (SC) belongs to science and technology, as social sciences and arts and humanity fields may be less likely to have technological impact. These steps leave us with $3\,737\,751$ unique papers. We assign research fields of papers based on their SC. Papers with multiple SCs are counted multiple times as individual observations, and our final dataset consists of $5\,611\,286$ observations.

We measure technological impact of our cohort of papers using citations received from utility patents granted at the USPTO between 1976 and 2012. We use a dataset that matched their front-page non-patent references to MEDLINE, so that for each and every non-patent reference, we know if and which MEDLINE paper it refers to \citep{Ke-compare-2018}. The matched results have an accuracy of 97\%.

\subsection{Dependent variables}

\begin{table*}
\centering 
\caption{Summary statistics of variables}
\label{tab:var}
\begin{tabular}{l r r r r r}
\hline
Variable & Mean & Std. Dev. & Min & Max & N \\ \hline
cited by patent & 0.112 & 0.315 & 0 & 1 & $5\,611\,286$ \\
\# patent citation & 0.510 & 4.407 & 0 & $2\,739$ & $5\,611\,286$ \\
patent citation lag & 9.550 & 5.457 & 0 & 32 & $628\,168$ \\
\# years to maximum patent citation & 1.128 & 2.869 & 0 & 30 & $628\,168$ \\
\# citing IPC & 2.784 & 2.144 & 0 & 76 & $628\,168$ \\
\hline
year & 1990.826 & 5.746 & 1980 & 1999 & $5\,611\,286$ \\
basicness & 1.024 & 0.894 & 0 & 2 & $5\,611\,286$ \\
NoveltyCat & 0.632 & 0.579 & 0 & 2 & $5\,611\,286$ \\
scientific citation & 31.795 & 95.645 & 0 & $59\,330$ & $5\,611\,286$ \\
journal Impact Factor & 2.098 & 2.52 & 0.001 & 39.104 & $5\,473\,178$ \\
international collaboration & 0.1 & 0.3 & 0 & 1 & $4\,358\,488$ \\
\# MeSH terms & 12.212 & 4.139 & 2 & 57 & $5\,611\,286$ \\
\# references & 19.453 & 14.103 & 0 & $1\,698$ & $5\,611\,286$ \\
\# authors & 4.073 & 2.608 & 1 & 546 & $5\,611\,286$ \\
\hline
\end{tabular}
\end{table*}

We introduce several dependent variables in our attempt to capture various dimensions of technological impact of scientific publications. The first one is whether the focal paper has achieved technological impact, that is, whether it has been cited by patents granted up until 2012. For the subset of papers with technological impact, we further consider the following variables that quantify various elements of technological impact:
\begin{itemize}
\item Intensity of impact, as measured by the number of citing patents;
\item Time lag of impact, as defined as the number of years passed from the publication year to the year when first patent citation occurred;
\item Longevity of impact, as quantified by the number of years from the year when the first citation occurred to the year when the paper got the most citations;
\item Scope of impact, as operationalized as the number of unique IPC (International Patent Classification) classes at the four-character level\footnote{E.g., ``C07D'' represents Heterocyclic Compounds.}.
\end{itemize}

Moreover, since patents at the USPTO are originated from many countries, although mainly the US, and assigned to diverse types of organizations (e.g., universities and companies), for robustness check, we restrict citing patents that are (1) from the US and (2) from US companies, and for both cases, recalculate all the introduced dependent variables.

\subsection{Independent variables}

Two independent variables of interest are basicness and novelty. Both of them are measured based on MeSH terms, which are controlled vocabularies used for indexing and retrieving MEDLINE papers. The list of MeSH terms of a paper is assigned by trained librarians at the National Library of Medicine (NLM) rather than authors, therefore is not subject to manipulation.

Basic and clinical research are the two broad types of science conducted within biomedical research. Clinical research is defined as ``research with human subjects''\footnote{https://grants.nih.gov/grants/glossary.htm\#ClinicalResearch}, whereas basic research advances understanding of ``living systems and life processes''\footnote{https://www.nigms.nih.gov/Education/Pages/factsheet\_CuriosityCreatesCures.aspx}. As such, basic research is conducted mostly at the level of cells and non-human animals, oftentimes using model organisms such as \emph{Caenorhabditis elegans} and \emph{Saccharomyces cerevisiae}. Therefore, one widely used criteria to distinguish whether a paper is basic science or clinical research is to check whether its associated MeSH terms have cell-, animal-, and human-related terms \citep{Ke-identify-2019}. Based on this, we introduce a categorical variable to indicate three levels of basicness: (1) clinical research (basicness = 0), if the MeSH terms of the focal paper contain only human-related terms; (2) moderately basic (basicness = 1), if both human and cell/animal terms are present in the MeSH terms; and (3) highly basic (basicness = 2), if there are only cell/animal terms.

In terms of novelty, we follow the literature that assesses novelty from the combinatorial perspective. In particular, similar to \citet{Boudreau-look-2016}, we compare pairwise combinations of MeSH terms with the existing literature. We first calculate the novelty score of the focal paper by looking at all the $n(n-1)/2$ pairs of MeSH terms, with $n$ denoting the number of terms, and counting the fraction of MeSH pairs that have not been combined in the entire previous literature, as captured by the MEDLINE database. We then construct our novelty measure as a categorical variable, which has three levels: (1) non-novel (NoveltyCat = 0), if the focal paper makes no new MeSH combination (novelty score 0); (2) moderately novel (NoveltyCat = 1), if the paper has at least one new combination (novelty score $>$ 0) but the novelty score is smaller than the top 5\% among papers clustered by subject category and year; and (3) highly novel (NoveltyCat = 2), if the novelty score is among the top 5\%.

\begin{table*}
\footnotesize
\centering
\caption{Fraction of papers that are cited by patents at the USPTO.}
\label{tab:overall}
\begin{tabular}{l | r | r r r | r r r r}
                               &         & \multicolumn{3}{c|}{Basicness} & \multicolumn{3}{c}{NoveltyCat} \\
                               & Overall & 0 & 1 & 2 & 0 & 1 & 2 \\
\hline
\# papers & $5\,611\,286$ & $2\,177\,939$ & $1\,121\,093$ & $2\,312\,254$ & $2\,352\,488$ & $2\,969\,692$ & $289\,106$ \\
\% cited by patents     & 11.19 & 4.54 & 18.76 & 13.80 & 7.94 & 13.50 & 13.96 \\
\% cited by US patents  & 8.84  & 3.64 & 15.04 & 10.75 & 6.20 & 10.71 & 11.21 \\
\% cited by US company patents & 5.66  & 2.38  & 9.52  & 6.87 & 3.94  & 6.85  & 7.44 \\
\end{tabular}
\end{table*}

\subsection{Control variables}

We use regression modeling techniques to understand how basicness and novelty are associated with technological impact, while controlling for other potential confounding factors that may have effects on patent citations as well as basicness and novelty. In light of previous studies, we take the following control variables into consideration.

The first one is the number of scientific citations---the number of scientific papers that cite the focal paper. Here we strict citing papers to those that were published up to 2012, since we count patent citations until that year. We consider this variable because prior works have found that scientific impact is positively associated with both technological impact \citep{Popp-from-2017, Ke-compare-2018} and novelty \citep{Uzzi-atypical-2013}.

We control for the venue of the focal paper using the journal Impact Factor (IF), as journals with high IF have high prestige, which may increase the visibility and facilitate the diffusion of the paper to the technology sphere. 

We further include an indicator encoding whether a paper involves international collaboration, as prior studies have established its linkage to novelty \citep{Wagner-intl-2019}.

Other control variables are the number of MeSH terms, number of references, and number of authors. They are commonly considered confounders and shown to have relations to novelty \citep{Uzzi-atypical-2013, Chai-breakthrough-2019} and technological impact \citep{Popp-from-2017}.

Finally, we create dummy variables for the WoS Subject Category and publication year. Our estimations, therefore, capture within-field and within-year differences; that is, the effects of basicness and novelty on technological impact are compared within papers in the same research field and published in the same year.

Table~\ref{tab:var} reports the summary statistics of our variables.

\section{Results}
\label{sec:res}

\subsection{Direct technological impact}

Table~\ref{tab:overall} shows that the characterization of papers as basic science links to higher probabilities of achieving technological impact, relative to clinical research. Overall, 11.2\% of papers are cited by patents that are granted until 2012. While 4.5\% of clinical research papers (basicness = 0) have technological impact, 13.8\% of highly basic papers---those with only cell/animal MeSH terms---do so. Interestingly, the group of moderately basic papers (basicness = 1) has the highest (18.8\%) fraction with technological impact. This means that papers that exhibit both the basic science component and the clinical research component are most likely to be listed as prior art by patented inventions. The same trend is observed if we limit citing patents to those from US organizations or from US companies.

As for novelty, as indicated from Table~\ref{tab:overall}, published biomedical literature displays low level of novelty, with 41.9\% of papers not introducing any new pairs of MeSH terms. This is consistent with previous studies \citep{Rzhetsky-choosing-2015, Foster-tradition-2015}. Table~\ref{tab:overall} reports that 7.9\% of these non-novel papers are cited by patents, but the fraction is much higher for novel papers. For the group of moderately novel papers (NoveltyCat = 1), a much higher fraction (13.5\%) of papers get cited by patents, and for the highly novel group, 14\% of papers are cited by patents. Again, this trend is robust if we consider only US patents or US company patents.

Next, we perform a series of logistic regressions to study the effects of basicness and novelty on technological impact, while controlling for the research field and publication year effects and other potential confounding factors as described before. In all models, the dependent variable is whether a paper has been directly cited by patents granted up to 2012. Table~\ref{tab:citedbypat} presents the modeling results, confirming positive associations found from Table~\ref{tab:overall}. While Model 1 in Table~\ref{tab:citedbypat} is the baseline where we included only control variables, Models 2--5 add independent variables. The reductions of BIC from Model 1 provide very strong support for Models 2--5.

\begin{table*}
\centering
\caption{Logistic regression results of whether a paper has been directly cited by USPTO patents granted until 2012.}
\label{tab:citedbypat}
\begin{tabular}{l*{5}{c}}
\hline\hline
                    &\multicolumn{1}{c}{(1)}&\multicolumn{1}{c}{(2)}&\multicolumn{1}{c}{(3)}&\multicolumn{1}{c}{(4)}&\multicolumn{1}{c}{(5)}\\
\hline
scientific citation ($\ln$)& 0.733\sym{***}& 0.751\sym{***}& 0.733\sym{***}& 0.750\sym{***}& 0.751\sym{***}\\
                           & (0.00159)     & (0.00161)     & (0.00159)     & (0.00161)     & (0.00161)     \\
[0.9em]
Impact Factor       &      0.0416\sym{***}&      0.0350\sym{***}&      0.0412\sym{***}&      0.0350\sym{***}&      0.0349\sym{***}\\
                    &  (0.000596)         &  (0.000592)         &  (0.000597)         &  (0.000593)         &  (0.000593)         \\
[0.9em]
\# MeSH             &      0.0130\sym{***}&     0.00731\sym{***}&     0.00407\sym{***}&   0.0000565         &    0.000257         \\
                    &  (0.000378)         &  (0.000383)         &  (0.000401)         &  (0.000406)         &  (0.000406)         \\
[0.9em]
\# ref ($\ln$)      &      0.0407\sym{***}&     -0.0310\sym{***}&      0.0449\sym{***}&     -0.0241\sym{***}&     -0.0240\sym{***}\\
                    &   (0.00239)         &   (0.00242)         &   (0.00240)         &   (0.00243)         &   (0.00243)         \\
[0.9em]
\# author ($\ln$)   &       0.236\sym{***}&       0.241\sym{***}&       0.235\sym{***}&       0.239\sym{***}&       0.239\sym{***}\\
                    &   (0.00295)         &   (0.00300)         &   (0.00295)         &   (0.00300)         &   (0.00300)         \\
[0.9em]
basicness=1         &                     &       0.904\sym{***}&                     &       0.873\sym{***}&       0.918\sym{***}\\
                    &                     &   (0.00507)         &                     &   (0.00508)         &   (0.00770)         \\
[0.9em]
basicness=2         &                     &       0.714\sym{***}&                     &       0.679\sym{***}&       0.676\sym{***}\\
                    &                     &   (0.00502)         &                     &   (0.00504)         &   (0.00705)         \\
[0.9em]
NoveltyCat=1        &                     &                     &       0.249\sym{***}&       0.204\sym{***}&       0.196\sym{***}\\
                    &                     &                     &   (0.00346)         &   (0.00349)         &   (0.00715)         \\
[0.9em]
NoveltyCat=2        &                     &                     &       0.499\sym{***}&       0.423\sym{***}&       0.663\sym{***}\\
                    &                     &                     &   (0.00666)         &   (0.00670)         &    (0.0153)         \\
[0.9em]
basicness=1 $\times$ NoveltyCat=1&        &                     &                     &                     &     -0.0328\sym{***}\\
                    &                     &                     &                     &                     &   (0.00938)         \\
[0.9em]
basicness=1 $\times$ NoveltyCat=2&        &                     &                     &                     &      -0.396\sym{***}\\
                    &                     &                     &                     &                     &    (0.0199)         \\
[0.9em]
basicness=2 $\times$ NoveltyCat=1&        &                     &                     &                     &      0.0271\sym{**} \\
                    &                     &                     &                     &                     &   (0.00846)         \\
[0.9em]
basicness=2 $\times$ NoveltyCat=2&        &                     &                     &                     &      -0.234\sym{***}\\
                    &                     &                     &                     &                     &    (0.0178)         \\
[0.9em]
Constant            &      -6.434\sym{***}&      -6.512\sym{***}&      -6.501\sym{***}&      -6.562\sym{***}&      -6.575\sym{***}\\
                    &    (0.0180)         &    (0.0181)         &    (0.0180)         &    (0.0181)         &    (0.0185)         \\
\hline
Field fe            & $\checkmark$        & $\checkmark$        & $\checkmark$        & $\checkmark$        & $\checkmark$        \\
Year fe             & $\checkmark$        & $\checkmark$        & $\checkmark$        & $\checkmark$        & $\checkmark$        \\
Observations        &     5473178         &     5473178         &     5473178         &     5473178         &     5473178         \\
Pseudo \(R^{2}\)    &       0.198         &       0.207         &       0.200         &       0.208         &       0.209         \\
\emph{BIC}          &   3089148           &   3055890           &   3081097           &   3050394           &   3050022           \\
\hline\hline
\multicolumn{6}{l}{\footnotesize Standard errors in parentheses}\\
\multicolumn{6}{l}{\footnotesize \sym{*} \(p<0.05\), \sym{**} \(p<0.01\), \sym{***} \(p<0.001\)}\\
\end{tabular}
\end{table*}

Let us elaborate the modeling results. Model 2 only examines the effect of basicness. We find that basic science is more likely to have technological impact than clinical research, and the size of such positive effect is large. As indicated from Model 2, after controlling for confounders, papers with intermediate level of basicness (basicness = 1) have $e^{0.904} - 1 = 147\%$ higher odds of being cited by patents than the odds for comparable clinical research papers in the same field and published in the year. For highly basic papers, there is 104\% increase in the odds of gaining technological impact. That moderately basic papers have a more pronounced effect on technological impact than highly basic papers indicates that papers that are most likely to get patent citations is those that have both basic science and clinical medicine components and resonates with the argument made by \citet{Gittelman-revolution-2016} that it is necessary for basic research to embed into clinical science to advance the understanding of the biological basis of diseases.

Model 3, which focuses on the novelty aspect only, indicates a significant positive effect of novelty on technological impact. After controlling for confounders, we expect 28.3\% and 64.7\% increase in the odds of receiving patent citations respectively for moderately and highly novel papers, compared to non-novel papers in the same field and published in the same year.

Models 4 considers both basicness and novelty, reassuring their positive influence on technological impact, regardless of whether or not controlling for each other. The comparison of Model 4 to Model 2 indicates that the coefficients of basicness do not decrease much after additionally controlling for novelty. On the other hand, comparing Model 4 to Model 3 indicates that additionally controlling for basicness decrease the positive coefficient of novelty.

\begin{figure*}
\centering
\includegraphics[trim=0mm 0mm 0mm 0mm, width=0.49\columnwidth]{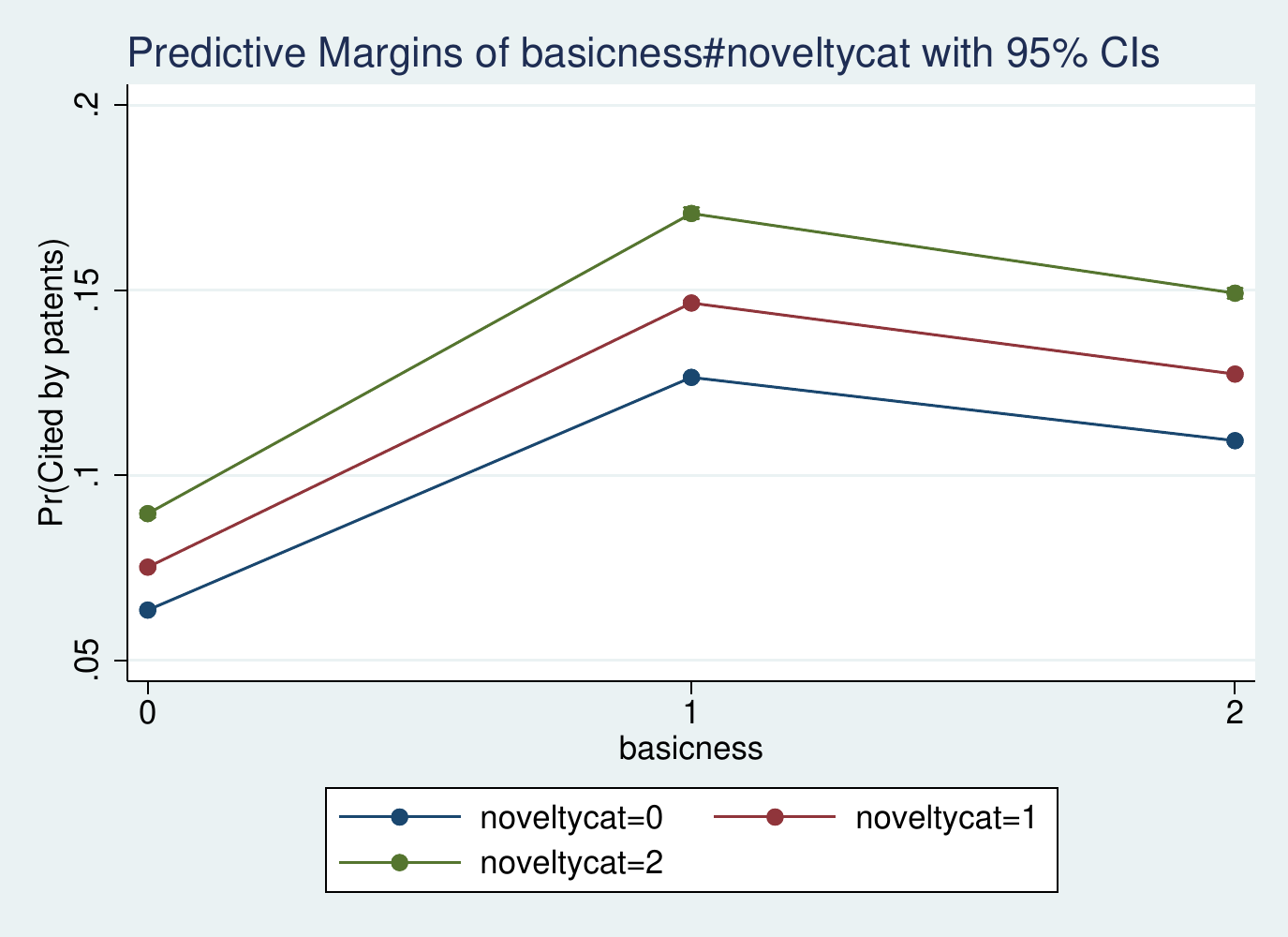}
\includegraphics[trim=0mm 0mm 0mm 0mm, width=0.49\columnwidth]{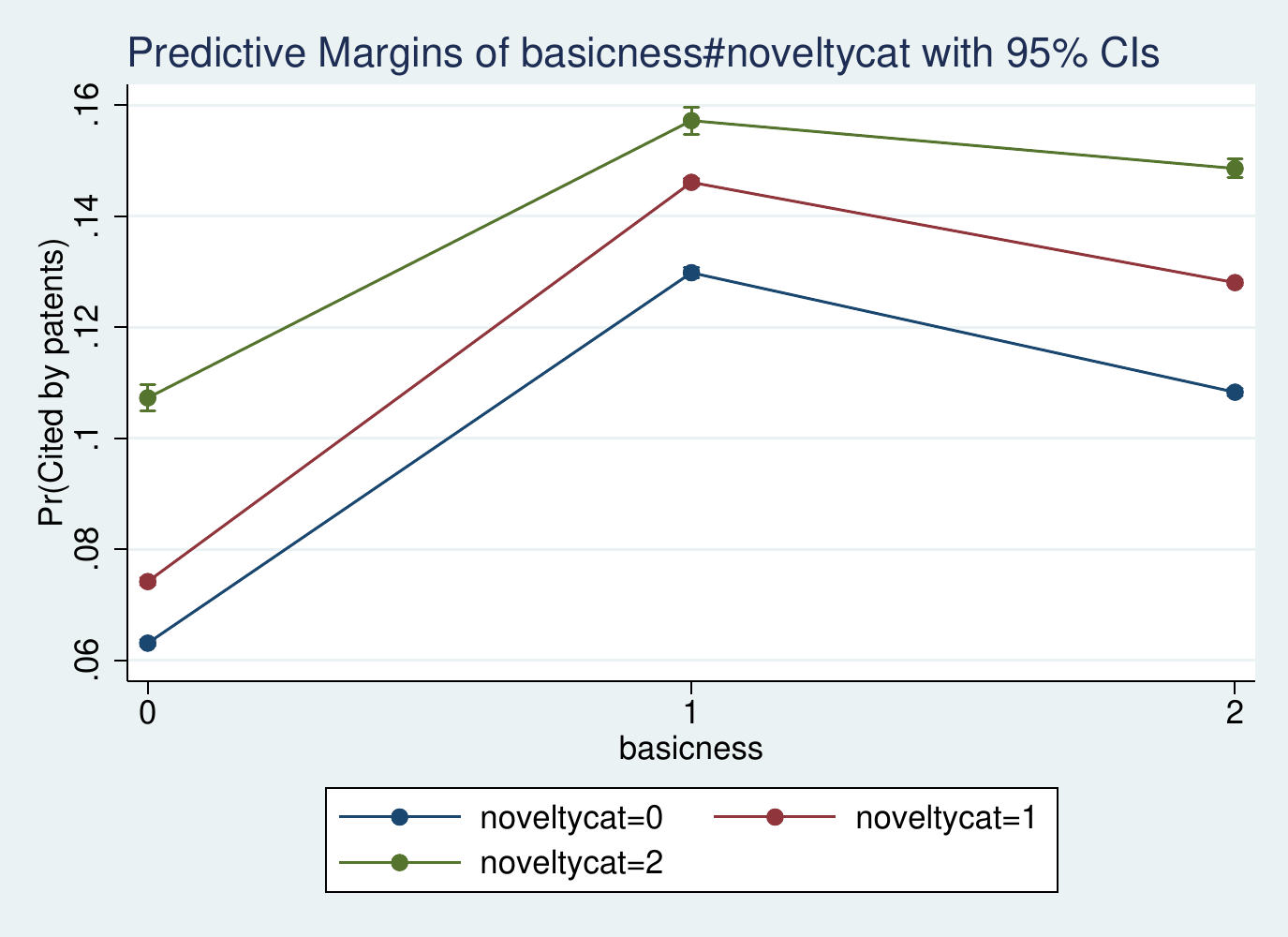}
\caption{Predictive margins of combinations of basicness and novelty (left) without and (right) with the interaction term.}
\label{fig:pm}
\end{figure*}

To further understand the effects of basicness and novelty, in the left panel in Fig.~\ref{fig:pm}, we show the predictive margins of the 9 combinations of different values of basicness and novelty. The predictive margin of a particular combination, for example moderately basic and moderately novel (basicness = 1, NoveltyCat = 1), is the average probability over all observations, where the probability of each observation is computed by setting its basicness and NoveltyCat values to 1 and fixing all other independent variable values as observed, \emph{i.e.}, treating it as though it were moderately basic and moderately novel, regardless of what its actual basicness and novelty are. We can see that papers that are simultaneously moderately basic and highly novel are most likely to obtain patent citations. We also observe that at each level of basicness, highly novel papers have larger probabilities to get patent citations than moderately novel papers, which are more likely to get cited than non-novel papers. The extent of differences between these margins is not equal, suggesting that the margin effect of novelty is dependent on basicness. This can be readily seen from Table~\ref{tab:ame} (the third and fourth columns), where we present the average marginal effects (AME) of novelty. AME for being moderately novel is 1.6\%; that is, on average a moderately novel paper's probability of being cited by patents is 1.6 percentage points higher than it is for non-novel novel. The effect differs greatly at varying levels of basicness; less than 1.2 percentage point for clinical papers and 1.8 percentage point for highly basic papers. A similar conclusion can be drawn for highly novel papers. On average, they have a 3.6 percentage point higher probability than non-novel papers, which decreases to 2.6 for clinical papers and increases to 4.4 for moderately basic papers.

Model 5 adds the basicness-novelty interaction term to Model 4. Again, for ease of interpretation, we plot the predictive margins in the right panel in Fig.~\ref{fig:pm}. The results in general agree with the model without the interaction term; that is, the combination of moderately basic and highly novel has the most pronounced effect on getting patent citations, novel papers have largest probabilities to get cited by patents at each level of basicness, and the differences of probabilities among different categories of novelty depend on levels of basicness (the last two columns in Table~\ref{tab:ame}).

Table~\ref{tab:citedbypat} also reveals a positive association between scientific impact and technological impact, in consistence with previous literature \citep{Ke-compare-2018, Popp-from-2017}. The positive linkage also holds for journal Impact Factor, indicating that papers published in more prestigious journals are more likely to be cited by patents than comparable papers.

\begin{table*}
\centering
\caption{Average marginal effects (AME) of novelty, varied by basicness and whether including the interaction term. All AMEs are statistically significant, with $z$ greater than 50 for the model without interaction and greater than 20 for the model with interaction.}
\label{tab:ame}
\begin{tabular}{c c || c c || c c}
\hline
NoveltyCat & Basicness & AME    & 95\% CI        & AME     & 95\% CI \\
\hline
1          & --        & 0.0164 & 0.0158  0.0169 & 0.0161  & 0.0155  0.0166 \\
1          & 0         & 0.0116 & 0.0112  0.0120 & 0.0111  & 0.0103  0.0119 \\
1          & 1         & 0.0201 & 0.0194  0.0208 & 0.0163  & 0.0151  0.0175 \\ 
1          & 2         & 0.0180 & 0.0174  0.0186 & 0.0197  & 0.0189  0.0205 \\
\hline
2          & --        & 0.0362 & 0.0350  0.0375 & 0.0374  & 0.0362  0.0387 \\
2          & 0         & 0.0261 & 0.0251  0.0270 & 0.0442  & 0.0418  0.0466 \\
2          & 1         & 0.0443 & 0.0428  0.0458 & 0.0274  & 0.0247  0.0301 \\
2          & 2         & 0.0398 & 0.0385  0.0412 & 0.0403  & 0.0385  0.0421 \\
\hline
\end{tabular}
\end{table*}

\subsection{Features of technological impact}

In the previous section, we have established that both basicness and novelty are positively associated with the likelihood to attain technological impact. We now restrict our analysis to the subsample of the $628\,168$ papers that are directly cited by patents and ask: Within this set of publications with technological impact, do the two characteristics still associate with various dimensions of technological impact? Here we examine four aspects, namely the intensity, time lag, longevity, and scope of the impact.

\begin{table*}
\centering
\caption{Characteristics of technological impact of papers that are cited by USPTO patents.}
\label{tab:tech}
\begin{tabular}{l | r | r r r | r r r }
                           &            & \multicolumn{3}{c|}{Basicness} & \multicolumn{3}{c}{NoveltyCat} \\
                           & Overall    & 0 & 1 & 2 & 0 & 1 & 2 \\
\hline
\# papers cited by patents & $628\,168$ & $98\,803$ & $210\,298$ & $319\,067$ & $186\,775$ & $401\,025$ & $40\,368$ \\
Average patent citations   & 4.55       & 3.60      & 4.97       & 4.57       & 4.12       & 4.68       & 5.33      \\
Average time lag           & 9.55       & 10.54     & 8.90       & 9.67       & 10.06      & 9.32       & 9.42      \\
Average duration           & 1.13       & 0.84      & 1.24       & 1.14       & 1.02       & 1.16       & 1.31      \\
Average \# of citing IPC   & 2.78       & 2.00      & 3.07       & 2.84       & 2.55       & 2.87       & 2.97      \\
\end{tabular}
\end{table*}

\subsubsection{Patent citation intensity}

Let us first focus on the intensity of technological impact, as measured by the number of patent citations. Similar to citations received from scientific papers, citation count from patents is also heavily distributed, as observed in \citet{Ke-compare-2018}. Within the studied sample, $254\,905$ ($40.6\%$) papers are cited by only one patent, but there is one paper that has attracted $2\,739$ patent citations.

Table~\ref{tab:tech} reports that on average clinical research papers have 3.6 patent citations, which is lower than citations of either moderately basic papers or highly basic papers. Novel science also possesses this superiority in achieving intense technological impact. While non-novel papers are cited by 4.1 patents, highly novel papers have 5.3 patent citations. We run negative binomial (NB) regression to examine whether the positive association persists after controlling for potential confounders. NB model is employed because patent citation is an over-dispersed variable. Model 2 in Table~\ref{tab:techimpact} presents the modeling results in the full model where all control variables are included, while Model 1 is the baseline model. We can confirm that both basic science and novel science are positively linked to the intensity of technological impact. Moderately and highly basic papers have respectively 14.9\% and 11.8\% more patent citations than comparable clinical papers. Moderately novel papers have 3.4\% more patent citations than non-novel papers, and the effect size is more pronounced for highly novel papers, reaching to 13.5\%.

\subsubsection{Patent citation time lag}

Next, we assess the time lag of technological impact, as defined as the number of years taken to get the first patent citation after the publication of the focal paper. Existing theories suggest that novel science faces resistance of acceptance from existing paradigms, and previous studies have empirically documented that novel science experiences delayed recognition in the scientific community. Given these works, it is important to ask whether delayed recognition for novel science also occurs in the technology community.

Table~\ref{tab:tech} indicates that this is not the case. Compared to clinical research papers that have 10.5 years of lag, it takes a smaller number of years for basic science papers to get the first patent citation. Similarly, novel science has a shorter lag than that of non-novel papers. We further use OLS to model the effects of basicness and novelty on time lag. Model 4 in Table~\ref{tab:techimpact} show negative associations, indicating that it requires less time for basic science papers and novel papers to get cited by patents than their respective counterparts. These results suggest that, on the contrary to what happens in the scientific domain, novel science appears not to display the delayed recognition phenomenon.

\begin{figure}
\centering
\includegraphics[trim=0mm 5mm 0mm 0mm, width=0.8\columnwidth]{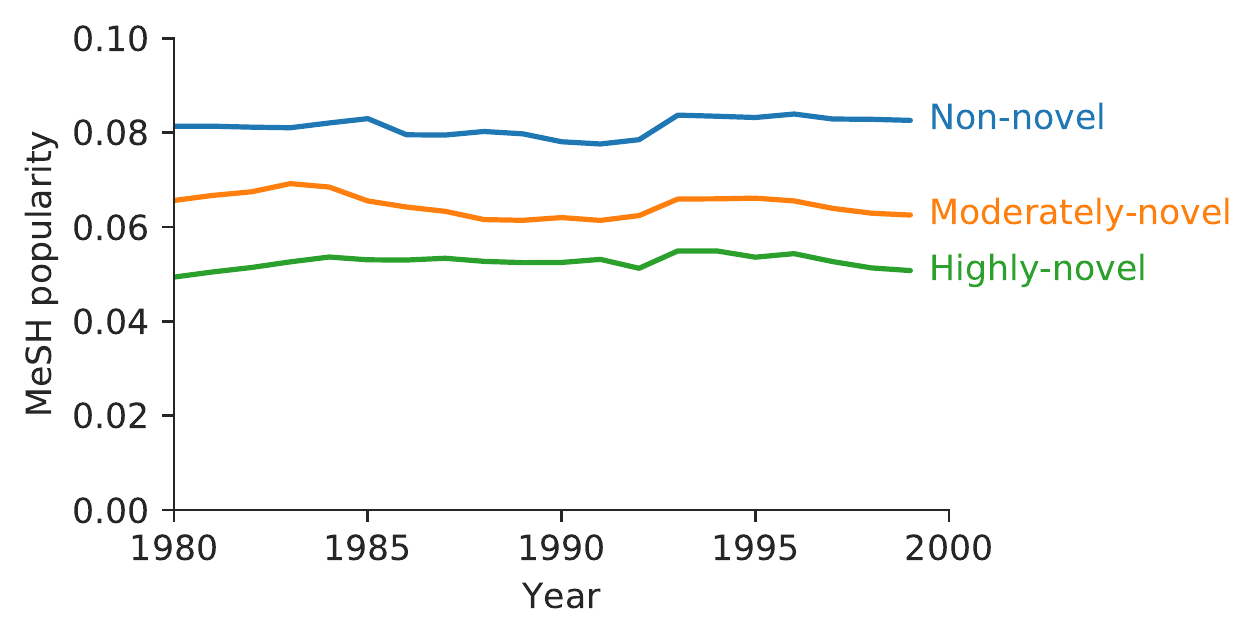}
\caption{MeSH term popularities of papers grouped by their novelty.}
\label{fig:popu}
\end{figure}

Why do novel science papers have a shorter time lag in getting the first patent citation than non-novel papers? Here we provide one explanation based on a recent work by \citet{Chai-breakthrough-2019}\footnote{We thank the reviewer for suggesting this.}. That is, novel papers cover less popular topics, in contrast to non-novel papers covering more popular topics, and this makes novel science face a less fierce competition for limited attention than non-novel papers, which facilitates the diffusion of novel science to the technology community. To provide evidence for this hypothesis, we first calculate the relative popularity of a MeSH term in a year, defined as the fraction of papers with that term among all papers published in that year. For each paper, we then take the average of the popularities of its MeSH terms as the indicator to measure to what extent it covers popular topics. Fig.~\ref{fig:popu} shows that for papers with technological impact, novel science covers less popular topics than non-novel science and this is consistent across years, supporting our hypothesis that the popularity of studied topics in a paper may have an effect on the time lag of technological impact.

\subsubsection{Patent citation longevity}

We also look at the other feature regarding the timing of technological impact, namely the longevity of impact, as quantified by the number of years taken to reach the maximum yearly patent citations from the year when the first citation occurred. Table~\ref{tab:tech} suggests that basic science papers have greater longevity than clinical research papers, so have novel papers than non-novel ones. OLS regression modeling assures that this positive influence of basicness and novelty on longevity holds after controlling for confounders, as demonstrated in Model 6 in Table~\ref{tab:techimpact}. 

\subsubsection{Patent citation scope}

Finally, we examine if basic science and novel science's technological impact spans broader technological domains, as operationalized by the number of 4-character IPC classes of their citing patents. Table~\ref{tab:tech} shows that this is indeed the case. While a clinical paper's citing patents on average belong to 2 classes, moderately basic and highly basic papers' citing patents respectively come from 3.1 and 2.8 classes. As for novelty, non-novel papers are cited by 2.5 IPC classes, in contrast to 3 classes for the highly novel papers. Further OLS modeling confirms positive associations between the scope of technological impact and basicness as well as novelty (Model 8 in Table~\ref{tab:techimpact}). Note that here we consider not only the complete set of control variables as before but also the number of patent citations as another confounder, because papers with a larger number of citing patents would by chance have more citing technological classes. Moderately basic papers have 0.33 more citing class than clinical papers, and 0.24 more for highly basic papers. Moderately and highly novel papers have respectively 0.03 and 0.1 more citing class than comparable non-novel papers.

\newgeometry{margin=1.3cm}
\begin{landscape}
\begin{table*}
\centering
\caption{Results of regression modeling of technological impact.}
\label{tab:techimpact}
\begin{tabular}{l*{8}{c}}
\hline\hline
                 &\multicolumn{2}{c}{Intensity (NB)}&\multicolumn{2}{c}{Time lag (OLS)}&\multicolumn{2}{c}{Longevity (OLS)}&\multicolumn{2}{c}{Scope (OLS)}\\
                 &\multicolumn{1}{c}{(1)}&\multicolumn{1}{c}{(2)}&\multicolumn{1}{c}{(3)}&\multicolumn{1}{c}{(4)}&\multicolumn{1}{c}{(5)}&\multicolumn{1}{c}{(6)}&\multicolumn{1}{c}{(7)}&\multicolumn{1}{c}{(8)}\\
\hline
scientific citation ($\ln$)& 0.281\sym{***} & 0.281\sym{***} & -0.362\sym{***} & -0.369\sym{***} & 0.446\sym{***} & 0.449\sym{***} & 0.109\sym{***} & 0.113\sym{***}\\
                           & (0.00122)      & (0.00123)      & (0.00622)       & (0.00623)       & (0.00352)      & (0.00353)      & (0.00177)      & (0.00177)     \\
[0.9em]
Impact Factor    & 0.0109\sym{***}  & 0.0105\sym{***}  & -0.0531\sym{***} & -0.0516\sym{***}& 0.0133\sym{***}  & 0.0127\sym{***}  & 0.0194\sym{***} & 0.0189\sym{***}\\
                 & (0.000385)       & (0.000384)       & (0.00191)        & (0.00191)       & (0.00108)        & (0.00108)        & (0.000532)      & (0.000532)     \\
[0.9em]
\# MeSH          & -0.00380\sym{***}& -0.00600\sym{***}& -0.00751\sym{***}& 0.00648\sym{***}& -0.00354\sym{***}& -0.00727\sym{***}& 0.00467\sym{***}& 0.000954\sym{*}\\
                 &  (0.000326)      &  (0.000349)      &   (0.00156)      &   (0.00167)     &  (0.000885)      &  (0.000945)      &  (0.000435)     &  (0.000464)    \\
[0.9em]
\# ref ($\ln$)   & -0.122\sym{***}  & -0.129\sym{***}  & 0.120\sym{***}   & 0.162\sym{***}  & -0.145\sym{***}  & -0.159\sym{***}  & 0.0263\sym{***} & 0.00667\sym{*} \\
                 & (0.00229)        & (0.00232)        & (0.0107)         & (0.0109)        & (0.00607)        & (0.00615)        & (0.00299)       & (0.00302)      \\
[0.9em]
\# author ($\ln$)& 0.0730\sym{***}  & 0.0707\sym{***}  & -0.394\sym{***}  & -0.381\sym{***} & 0.143\sym{***}   & 0.139\sym{***}   & 0.0326\sym{***} & 0.0274\sym{***} \\
                 & (0.00254)        & (0.00255)        & (0.0124)         & (0.0124)        & (0.00700)        & (0.00703)        & (0.00344)       & (0.00346)       \\
[0.9em]
basicness=1      &                  & 0.139\sym{***}   &                  & -0.754\sym{***} &                  & 0.246\sym{***}   &                 & 0.326\sym{***}\\
                 &                  &   (0.00490)      &                  & (0.0227)        &                  & (0.0128)         &                 & (0.00630)     \\
[0.9em]
basicness=2      &                  & 0.112\sym{***}   &                  & -0.554\sym{***} &                  & 0.187\sym{***}   &                 & 0.237\sym{***}\\
                 &                  & (0.00471)        &                  & (0.0219)        &                  &    (0.0124)      &                 &   (0.00610)   \\
[0.9em]
NoveltyCat=1     &                  & 0.0331\sym{***}  &                  & -0.218\sym{***} &                  & 0.0513\sym{***}  &                 & 0.0285\sym{***}\\
                 &                  &   (0.00320)      &                  &    (0.0151)     &                  &   (0.00855)      &                 &   (0.00420)    \\
[0.9em]
NoveltyCat=2     &                  & 0.127\sym{***}   &                  &  -0.494\sym{***}&                  &    0.200\sym{***}&                 & 0.0997\sym{***}\\
                 &                  &   (0.00578)      &                  &    (0.0278)     &                  &    (0.0157)      &                 &   (0.00771)    \\
[0.9em]
patent citation ($\ln$)&            &                  &                  &                 &                  &                  &   1.579\sym{***}&  1.574\sym{***}\\
                 &                  &                  &                  &                 &                  &                  &   (0.00193)     &   (0.00193)    \\
[0.9em]
Constant         &    0.268\sym{***}&    0.235\sym{***}&    18.08\sym{***}&   18.25\sym{***}&   -0.496\sym{***}&   -0.555\sym{***}&   0.353\sym{***}&  0.302\sym{***}\\
                 &    (0.0172)      &    (0.0172)      &    (0.0811)      &    (0.0812)     &    (0.0458)      &    (0.0460)      &    (0.0226)     &    (0.0226)    \\
\hline
lnalpha          &   -0.204\sym{***}&   -0.207\sym{***}&                  &                 &                  &                  &                 &                \\
                 &   (0.00210)      &   (0.00210)      &                  &                 &                  &                  &                 &                \\
\hline
Field fe         & $\checkmark$ & $\checkmark$ & $\checkmark$ & $\checkmark$ & $\checkmark$ & $\checkmark$ & $\checkmark$ & $\checkmark$ \\
Year fe          & $\checkmark$ & $\checkmark$ & $\checkmark$ & $\checkmark$ & $\checkmark$ & $\checkmark$ & $\checkmark$ & $\checkmark$ \\
Observations     & 615981       & 615981       & 615981       & 615981       & 615981       & 615981       & 615981       & 615981       \\
(Pseudo) \(R^{2}\)& 0.0384      & 0.0389       & 0.179        & 0.181        & 0.0526       & 0.0535       & 0.588        & 0.590        \\
\textit{BIC}     & 3104621      & 3103322      & 3719125      & 3717606      & 3016986      & 3016484      & 2142837      & 2139976      \\
\hline\hline
\multicolumn{9}{l}{\footnotesize Standard errors in parentheses. NB: negative binomial.}\\
\multicolumn{9}{l}{\footnotesize \sym{*} \(p<0.05\), \sym{**} \(p<0.01\), \sym{***} \(p<0.001\)}\\
\end{tabular}
\end{table*}
\end{landscape}
\restoregeometry

\subsection{Robustness tests}

We perform several additional regressions to test the robustness of our findings. First, in all the analyses presented as far, we have not taken into account the variable whether a paper involves international collaboration, because there is a large fraction (22\%) of papers without enough affiliation data that allows us to infer the variable (Table~\ref{tab:var}). This is especially the case for papers in earlier years. Our results, nevertheless, remain robust if we restrict our sample to those papers with known value for this variable and included it as another control variable (Tables~\ref{tab:citedbypat:intl} and \ref{tab:techimpact:intl}). Interestingly, international collaboration has a negative effect on achieving technological impact, in contrast to its positive effect in the scientific community where papers involving international collaboration have more scientific citations.

Patents may originate from varying types of organizations. Although the vast majority of patents are assigned to companies, public research organizations still own patents, especially in our focused sector of life and biomedical science. As the contribution to private sector innovations is of particular interest to many policymakers, when measuring technological impact, we only considered citing patents (1) from US organizations and from (2) US companies. We then repeat all analyses and find that all the findings hold (Tables~\ref{tab:citedbypatus} and \ref{tab:techimpact:us}; Tables~\ref{tab:citedbypatuscom} and \ref{tab:techimpact:uscom}).

We have used negative binomial regression to model the intensity of patent citations for papers with technological impact. To further examine the robustness of our results, we have performed two checks: (1) OLS with $\log_{10}$-transformed number of patent citations as the dependent variable, and (2) logit models where the dependent variables are whether the number of patent citations is ranked into the top 5\% (or top 10\%) for papers in the same field and the same year. Both checks consolidate our findings that basic science and novel science have positive effects on number of patent citations (Table~\ref{tab:patc}).

Finally, we have duplicated papers with multiple SCs as individual observations. We adopted an alternative way, whereby we chose from a paper's associated SCs the one that is most presented in its reference list. This practice has also been used in previous literature (e.g., \citet{Verhoeven-measuring-2016}). Our findings still stand (Tables~\ref{tab:citedbypat:ssc} and \ref{tab:techimpact:ssc}).

\section{Concluding discussion}
\label{sec:dis}

In this work, we have studied science-technology interaction from the science perspective by identifying which characteristics of scientific publications are more likely to garner technological impact. We have paid attention to two important features, namely basicness and novelty. Drawing on a large-scale corpus consisting of 3.8 million MEDLINE research articles published between 1980 and 1999 and all the USPTO utility patents granted from 1976 to 2012, we have systematically investigated the relationships between technological impact and basicness and novelty. We found that both features have positive associations with technological impact and the effect sizes are large. In particular, our estimations indicate that, after controlling for confounders, the odds of being cited by patents is 139\% higher for moderately basic science papers than comparable clinical research papers, and 97\% higher for highly basic papers. Moderately and highly novel papers' odds of being cited by patents increase 23\% and 53\%, compared to non-novel papers. Such superiority in obtaining technological impact for basic science and novel publications is robust even after controlling for their scientific impact and the prestige of the journal where they were published, suggesting potential factors beyond scientific quality and venue visibility arising for the technological impact. Moreover, the positive effects are persistent across the study period.

We have further confined the sample to papers that are cited by patents and explored technological impact profiles of those papers. We found that both basic science papers and novel papers consistently have more patent citations and a broader scope of technological impact. Moreover, it takes less time for both of them to reach to the technology space. This relative absence of time delay of novel science is in contrast to what happens in the scientific community, where novel science has displayed a delayed recognition in gaining scientific impact.

Our work contributes to the ongoing understanding of the interaction between science and technology and contrasts to many earlier studies that examined the linkage from the technology perspective. First, we find that in biomedicine, basic research papers are more likely to get patent citations. Although the role of basic science in technology development has been discussed previously \citep{Narin-linkage-1997, McMillan-biotech-2000, Lim-relationship-2004}, our systematic study differs from them in several ways. Specifically, analyses presented in both \citet{Narin-linkage-1997} and \citet{McMillan-biotech-2000} were \emph{retrospective}; they started from patents and traced back to examine which types of papers get cited by patents. Ours, by contrast, is \emph{prospective}; our starting point is all published papers and we examine if basic science papers are more likely to get patent citations than clinical papers. Such a distinction is important, because patent-centered, retrospective studies fail to take population papers into account, and the observation made in \citet{Narin-linkage-1997} and \citet{McMillan-biotech-2000} that cited science is basic could simply due to the fact that there are more basic papers, as we have found previously \citep{Ke-identify-2019}, rather than their advantage in gaining technological impact. Our work is a direct demonstration of the advantage of basic science. Moreover, we quantified through regression techniques the effect of basicness on the likelihood of receiving patent citations and further explored different characteristics of technological impact. Both features were absent in \citet{Narin-linkage-1997} and \citet{McMillan-biotech-2000}. Our work asks a different research question from the one asked in \citet{Lim-relationship-2004}, which focused on firms and studied the relationship between their innovation output, as measured by number of patents, and within-firm basic/applied research activities, as measured by number of publications.

Second, we studied the combination of basicness and novelty and potential interactions between them. We found that papers featuring moderately basic and highly novel have largest probabilities of getting cited by patents and at each level of basicness, highly novel papers are most likely to accrue patent citations. Moreover, we presented a large-scale and systematic study of the role of novelty in technological impact; we looked at every research article indexed in the most comprehensive database for biomedical research literature in a 20-year period, as opposed to papers published in a single year. The scale of a corpus matters, because conclusions drawn from a large-scale corpus may point to general and robust features of the system rather than properties that are potentially constrained by peculiarities of a small-scale corpus. This aligns with recent interests in computational social science \citep{Lazer-css-2009} and science of science \citep{Fortunato-scisci-2018, Zeng-scisci-2017}.

Our findings speak to the ongoing discussions in the science policy literature on the role of basic science and novel science. Recently, there is an increasing pressure to demonstrate the societal impact of science. In biomedicine, basic biomedical research has received comments about its relevance in improving societal health, based on the superficial view that basic science oftentimes uses model organisms as study subjects. However, by using model organisms, basic science sheds light on life processes and disease mechanisms, and knowledge produced from basic science ``leads to better ways to predict, prevent, diagnose, and treat disease,'' therefore improving patient health. Moreover, in drug development, there is a linear pipeline where basic science must precede clinical research. Only after basic science with certain goals such as target and drug identification and preclinical validation has been fulfilled can the development enter the clinical trial and registration stage. Our work adds another dimension of the impact of basic science, that is, it contributes to technology development.

As for novelty, there have been increasing concerns that funding agencies are conservative in choosing projects, favoring safe ones and biasing against novel proposals that plan to chart less explored research topics \citep{Boudreau-look-2016}. Yet, prior studies suggested that funding policies are among the key drivers of scientific creativity \citep{Azoulay-incentives-2011}, and novel science is an important source of scientific breakthrough \citep{Uzzi-atypical-2013}. Our work aligns with these lines of inquiry and provides empirical, supporting evidence for alleviating the bias against novelty, or even encouraging the pursuit of novel science, because it contributes to the development of technological innovations.

We have looked at whether and when a scientific paper has technological impact as well as the intensity and scope of the impact, without assessing the characteristics of citing patents. Future work can explore whether basic and novel science are more likely to have technological impact from high quality, novel, or disruptive patents. Moreover, we have only considered the biomedical research area and patents at the USPTO, it remains to be seen if the advantage of achieving technological impact still holds for other scientific disciplines like physical sciences and engineering and for patents from other patent offices. In addition, uncovering more \emph{ex~ante} characteristics of scientific publications, such as interdisciplinarity, may also serve as another topic for future investigations. Finally, our work raises several questions related to the context and pathway from basic and novel science to technological innovations. For example, do author-inventor play a role in forming the linkage? Does geographic distance or science-technology knowledge relatedness matter more? Answers to these questions would further advance the understanding of science-technology linkage.

\section*{Acknowledgments}

I thank the anonymous referees for helpful comments and suggestions. This work used computing resources and \texttt{Stata} provided by Northeastern University.

\appendix

\setcounter{figure}{0}
\makeatletter 
\renewcommand{\thefigure}{A.\@arabic\c@figure}
\makeatother

\setcounter{table}{0}
\makeatletter 
\renewcommand{\thetable}{A.\@arabic\c@table}
\makeatother

\begin{table*}
\centering
\caption{Logit regression of direct patent citation. All models include whether a focal paper involves international collaboration as another control variable. The reductions of BIC provide very strong support for Models 2--5.}
\label{tab:citedbypat:intl}
\begin{tabular}{l*{5}{c}}
\hline\hline
                    &\multicolumn{1}{c}{(1)}&\multicolumn{1}{c}{(2)}&\multicolumn{1}{c}{(3)}&\multicolumn{1}{c}{(4)}&\multicolumn{1}{c}{(5)}\\
\hline
scientific citation ($\ln$)& 0.730\sym{***}& 0.750\sym{***}& 0.730\sym{***}& 0.749\sym{***}& 0.750\sym{***} \\
                           & (0.00178)     & (0.00180)     & (0.00178)     & (0.00180)     & (0.00181)      \\
[0.7em]
Impact Factor       &      0.0432\sym{***}&      0.0359\sym{***}&      0.0429\sym{***}&      0.0360\sym{***}&      0.0359\sym{***}\\
                    &  (0.000671)         &  (0.000665)         &  (0.000672)         &  (0.000666)         &  (0.000666)         \\
[0.7em]
\# MeSH             &      0.0142\sym{***}&     0.00874\sym{***}&     0.00560\sym{***}&     0.00191\sym{***}&     0.00212\sym{***}\\
                    &  (0.000412)         &  (0.000417)         &  (0.000437)         &  (0.000442)         &  (0.000442)         \\
[0.7em]
\# ref ($\ln$)      &      0.0374\sym{***}&     -0.0446\sym{***}&      0.0418\sym{***}&     -0.0374\sym{***}&     -0.0373\sym{***}\\
\end{tabular}
\end{table*}

\newgeometry{margin=1.3cm}
\begin{landscape}
\begin{table*}
\centering
\caption{Results of regression modeling of technological impact. All models include whether a focal paper involves international collaboration as another control variable.}
\label{tab:techimpact:intl}
\begin{tabular}{l*{8}{c}}
\hline\hline
                    & \multicolumn{2}{c}{Intensity (NB)} & \multicolumn{2}{c}{Time lag (OLS)} & \multicolumn{2}{c}{Longevity (OLS)} & \multicolumn{2}{c}{Scope (OLS)} \\
                    &\multicolumn{1}{c}{(1)}&\multicolumn{1}{c}{(2)}&\multicolumn{1}{c}{(3)}&\multicolumn{1}{c}{(4)}&\multicolumn{1}{c}{(5)}&\multicolumn{1}{c}{(6)}&\multicolumn{1}{c}{(7)}&\multicolumn{1}{c}{(8)}\\
\hline
scientific citation ($\ln$)& 0.280\sym{***}& 0.281\sym{***}& -0.361\sym{***}& -0.370\sym{***}& 0.444\sym{***}& 0.447\sym{***}& 0.109\sym{***}& 0.114\sym{***}\\
                           &   (0.00136)   &   (0.00136)   &   (0.00672)    &   (0.00673)    &   (0.00385)   &   (0.00386)   &   (0.00197)   &   (0.00197)   \\
[0.9em]
Impact Factor       & 0.0106\sym{***}  & 0.0103\sym{***}  & -0.0520\sym{***} & -0.0501\sym{***}& 0.0130\sym{***}& 0.0123\sym{***}& 0.0190\sym{***}& 0.0183\sym{***}\\
                    & (0.000426)       & (0.000425)       & (0.00206)        & (0.00206)       & (0.00118)      & (0.00118)      & (0.000591)     & (0.000591)     \\
[0.9em]
\# MeSH             & -0.00296\sym{***}& -0.00500\sym{***}& -0.00900\sym{***}& 0.00349\sym{*}  & -0.00190\sym{*}  & -0.00538\sym{***}& 0.00480\sym{***}& 0.00130\sym{**} \\
                    & (0.000351)       & (0.000376)       & (0.00164)        & (0.00175)         & (0.000942)         & (0.00101)         & (0.000471)         & (0.000502)         \\
\end{tabular}
\end{table*}
\end{landscape}
\restoregeometry

\begin{table*}
\centering
\caption{Logit regression of direct patent citation. The dependent variables are based on patents assigned to US organizations.}
\label{tab:citedbypatus}
\begin{tabular}{l*{5}{c}}
\hline\hline
                    &\multicolumn{1}{c}{(1)}&\multicolumn{1}{c}{(2)}&\multicolumn{1}{c}{(3)}&\multicolumn{1}{c}{(4)}&\multicolumn{1}{c}{(5)}\\
\hline
scientific citation ($\ln$)&       0.753\sym{***}&       0.769\sym{***}&       0.753\sym{***}&       0.768\sym{***}&       0.768\sym{***}\\
                           &   (0.00174)         &   (0.00176)         &   (0.00174)         &   (0.00176)         &   (0.00176)         \\
[0.9em]
Impact Factor       &      0.0421\sym{***}&      0.0368\sym{***}&      0.0418\sym{***}&      0.0369\sym{***}&      0.0368\sym{***}\\
                    &  (0.000608)         &  (0.000605)         &  (0.000609)         &  (0.000606)         &  (0.000606)         \\
[0.9em]
\# MeSH             &      0.0128\sym{***}&     0.00702\sym{***}&     0.00385\sym{***}&   -0.000266         &  -0.0000519         \\
                    &  (0.000413)         &  (0.000418)         &  (0.000438)         &  (0.000443)         &  (0.000443)         \\
[0.9em]
\# ref ($\ln$)      &      0.0344\sym{***}&     -0.0340\sym{***}&      0.0385\sym{***}&     -0.0271\sym{***}&     -0.0270\sym{***}\\
\end{tabular}
\end{table*}

\newgeometry{margin=1.3cm}
\begin{landscape}
\begin{table*}
\centering
\caption{Results of regression modeling of technological impact. The dependent variables are based on patents assigned to US organizations.}
\label{tab:techimpact:us}
\begin{tabular}{l*{8}{c}}
\hline\hline
                    & \multicolumn{2}{c}{Intensity (NB)} & \multicolumn{2}{c}{Time lag (OLS)} & \multicolumn{2}{c}{Longevity (OLS)} & \multicolumn{2}{c}{Scope (OLS)} \\
                    &\multicolumn{1}{c}{(1)}&\multicolumn{1}{c}{(2)}&\multicolumn{1}{c}{(3)}&\multicolumn{1}{c}{(4)}&\multicolumn{1}{c}{(5)}&\multicolumn{1}{c}{(6)}&\multicolumn{1}{c}{(7)}&\multicolumn{1}{c}{(8)}\\
\hline
scientific citation ($\ln$)&       0.282\sym{***}&       0.282\sym{***}&      -0.308\sym{***}&      -0.314\sym{***}&       0.387\sym{***}&       0.389\sym{***}&       0.107\sym{***}&       0.111\sym{***}\\
                           &   (0.00151)         &   (0.00151)         &   (0.00693)         &   (0.00694)         &   (0.00377)         &   (0.00377)         &   (0.00196)         &   (0.00196)         \\
[0.9em]
Impact Factor       &      0.0148\sym{***}&      0.0145\sym{***}&     -0.0520\sym{***}&     -0.0509\sym{***}&      0.0112\sym{***}&      0.0106\sym{***}&      0.0181\sym{***}&      0.0176\sym{***}\\
                    &  (0.000481)         &  (0.000481)         &   (0.00205)         &   (0.00205)         &   (0.00111)         &   (0.00111)         &  (0.000570)         &  (0.000569)         \\
[0.9em]
\# MeSH             &    -0.00461\sym{***}&    -0.00693\sym{***}&    -0.00813\sym{***}&     0.00517\sym{**} &    -0.00565\sym{***}&    -0.00879\sym{***}&     0.00488\sym{***}&     0.00101         \\
                    &  (0.000402)         &  (0.000431)         &   (0.00175)         &   (0.00186)         &  (0.000949)         &   (0.00101)         &  (0.000486)         &  (0.000518)         \\
\end{tabular}
\end{table*}
\end{landscape}
\restoregeometry

\begin{table*}
\centering
\caption{Logit regression of direct patent citation. The dependent variables are based on patents assigned to US companies.}
\label{tab:citedbypatuscom}
\begin{tabular}{l*{5}{c}}
\hline\hline
                    &\multicolumn{1}{c}{(1)}&\multicolumn{1}{c}{(2)}&\multicolumn{1}{c}{(3)}&\multicolumn{1}{c}{(4)}&\multicolumn{1}{c}{(5)}\\
\hline
scientific citation ($\ln$)&       0.734\sym{***}&       0.746\sym{***}&       0.734\sym{***}&       0.744\sym{***}&       0.745\sym{***}\\
                           &   (0.00204)         &   (0.00205)         &   (0.00204)         &   (0.00205)         &   (0.00205)         \\
[0.9em]
Impact Factor       &      0.0210\sym{***}&      0.0174\sym{***}&      0.0206\sym{***}&      0.0173\sym{***}&      0.0172\sym{***}\\
                    &  (0.000632)         &  (0.000635)         &  (0.000633)         &  (0.000636)         &  (0.000636)         \\
[0.9em]
\# MeSH             &      0.0113\sym{***}&     0.00622\sym{***}&     0.00292\sym{***}&   -0.000680         &   -0.000432         \\
                    &  (0.000490)         &  (0.000496)         &  (0.000520)         &  (0.000525)         &  (0.000525)         \\
[0.9em]
\# ref ($\ln$)      &     -0.0133\sym{***}&     -0.0764\sym{***}&    -0.00900\sym{**} &     -0.0693\sym{***}&     -0.0693\sym{***}\\
\end{tabular}
\end{table*}

\newgeometry{margin=1.3cm}
\begin{landscape}
\begin{table*}
\centering
\caption{Results of regression modeling of technological impact. The dependent variables are based on patents assigned to US companies.}
\label{tab:techimpact:uscom}
\begin{tabular}{l*{8}{c}}
\hline\hline
                    & \multicolumn{2}{c}{Intensity (NB)} & \multicolumn{2}{c}{Time lag (OLS)} & \multicolumn{2}{c}{Longevity (OLS)} & \multicolumn{2}{c}{Scope (OLS)} \\
                    &\multicolumn{1}{c}{(1)}&\multicolumn{1}{c}{(2)}&\multicolumn{1}{c}{(3)}&\multicolumn{1}{c}{(4)}&\multicolumn{1}{c}{(5)}&\multicolumn{1}{c}{(6)}&\multicolumn{1}{c}{(7)}&\multicolumn{1}{c}{(8)}\\
\hline
scientific citation ($\ln$)& 0.263\sym{***}& 0.263\sym{***}& -0.264\sym{***}& -0.270\sym{***}& 0.333\sym{***}& 0.334\sym{***}& 0.0911\sym{***}& 0.0947\sym{***}\\
                           & (0.00215)     & (0.00216)     & (0.00875)      & (0.00875)      & (0.00446)     & (0.00447)     & (0.00239)      & (0.00239)         \\
[0.9em]
Impact Factor       & 0.00790\sym{***} & 0.00764\sym{***}& -0.0360\sym{***}& -0.0344\sym{***}& 0.00562\sym{***}& 0.00524\sym{***}& 0.0136\sym{***}& 0.0129\sym{***}\\
                    &  (0.000701)      & (0.000701)      & (0.00265)       & (0.00265)       & (0.00135)       & (0.00135)       & (0.000716)     &  (0.000715)    \\
[0.9em]
\# MeSH             & -0.00670\sym{***}& -0.00832\sym{***}& -0.00338         & 0.00887\sym{***}& -0.00866\sym{***}& -0.0111\sym{***}& 0.00394\sym{***}& 0.0000832         \\
                    & (0.000581)       & (0.000624)         & (0.00225)         & (0.00240)         & (0.00115)         & (0.00123)         & (0.000607)         &  (0.000648)         \\
\end{tabular}
\end{table*}
\end{landscape}
\restoregeometry

\begin{table*}
\centering
\caption{Results of regression modeling of number of patent citations. The variance inflation factor statistics are about 2, indicating no multicollinearity among the independent variables.}
\label{tab:patc}
\begin{tabular}{l*{6}{c}}
\hline\hline
                    &\multicolumn{2}{c}{$\log_{10}$(patent citation) (OLS)} & \multicolumn{2}{c}{Top 5\% (logit)} & \multicolumn{2}{c}{Top 10\% (logit)} \\
                    &\multicolumn{1}{c}{(1)}&\multicolumn{1}{c}{(2)}&\multicolumn{1}{c}{(3)}&\multicolumn{1}{c}{(4)}&\multicolumn{1}{c}{(5)}&\multicolumn{1}{c}{(6)}\\
\hline
scientific citation ($\ln$)&      0.0861\sym{***}&      0.0866\sym{***}&       0.732\sym{***}&       0.734\sym{***}&       0.595\sym{***}&       0.599\sym{***}\\
                           &  (0.000496)         &  (0.000496)         &   (0.00614)         &   (0.00616)         &   (0.00450)         &   (0.00452)         \\
[0.9em]
Impact Factor       &     0.00495\sym{***}&     0.00482\sym{***}&     0.00951\sym{***}&     0.00904\sym{***}&     0.00955\sym{***}&     0.00891\sym{***}\\
                    &  (0.000152)         &  (0.000152)         &   (0.00126)         &   (0.00127)         &   (0.00101)         &   (0.00102)         \\
[0.9em]
\# MeSH             &   0.0000660         &   -0.000825\sym{***}&    -0.00535\sym{***}&     -0.0101\sym{***}&    -0.00229\sym{*}  &    -0.00691\sym{***}\\
                    &  (0.000125)         &  (0.000133)         &   (0.00141)         &   (0.00151)         &   (0.00103)         &   (0.00111)         \\
\end{tabular}
\end{table*}

\begin{table*}
\centering
\caption{Logit regression of direct patent citation. Papers with multiple SC are assigned to the one most presented in their reference list.}
\label{tab:citedbypat:ssc}
\begin{tabular}{l*{5}{c}}
\hline\hline
                    &\multicolumn{1}{c}{(1)}&\multicolumn{1}{c}{(2)}&\multicolumn{1}{c}{(3)}&\multicolumn{1}{c}{(4)}&\multicolumn{1}{c}{(5)}\\
\hline
scientific citation ($\ln$)&       0.758\sym{***}&       0.775\sym{***}&       0.758\sym{***}&       0.774\sym{***}&       0.774\sym{***}\\
                           &   (0.00199)         &   (0.00201)         &   (0.00199)         &   (0.00201)         &   (0.00201)         \\
[0.9em]
Impact Factor       &      0.0375\sym{***}&      0.0318\sym{***}&      0.0372\sym{***}&      0.0319\sym{***}&      0.0318\sym{***}\\
                    &  (0.000767)         &  (0.000763)         &  (0.000768)         &  (0.000765)         &  (0.000765)         \\
[0.9em]
\# MeSH             &      0.0133\sym{***}&     0.00755\sym{***}&     0.00427\sym{***}&   0.0000976         &    0.000328         \\
                    &  (0.000471)         &  (0.000476)         &  (0.000499)         &  (0.000504)         &  (0.000505)         \\
[0.9em]
\# ref ($\ln$)      &      0.0233\sym{***}&     -0.0397\sym{***}&      0.0271\sym{***}&     -0.0333\sym{***}&     -0.0332\sym{***}\\
\end{tabular}
\end{table*}

\newgeometry{margin=1.3cm}
\begin{landscape}
\begin{table*}
\centering
\caption{Results of regression modeling of technological impact. Papers with multiple SC are assigned to the one most presented in their reference list.}
\label{tab:techimpact:ssc}
\begin{tabular}{l*{8}{c}}
\hline\hline
                    & \multicolumn{2}{c}{Intensity (NB)} & \multicolumn{2}{c}{Time lag (OLS)} & \multicolumn{2}{c}{Longevity (OLS)} & \multicolumn{2}{c}{Scope (OLS)} \\
                    &\multicolumn{1}{c}{(1)}&\multicolumn{1}{c}{(2)}&\multicolumn{1}{c}{(3)}&\multicolumn{1}{c}{(4)}&\multicolumn{1}{c}{(5)}&\multicolumn{1}{c}{(6)}&\multicolumn{1}{c}{(7)}&\multicolumn{1}{c}{(8)}\\
\hline
scientific citation ($\ln$)&       0.290\sym{***}&       0.291\sym{***}&      -0.399\sym{***}&      -0.405\sym{***}&       0.465\sym{***}&       0.467\sym{***}&       0.110\sym{***}&       0.114\sym{***}\\
                           &   (0.00152)         &   (0.00152)         &   (0.00773)         &   (0.00774)         &   (0.00443)         &   (0.00444)         &   (0.00222)         &   (0.00222)         \\
[0.9em]
Impact Factor       &     0.00978\sym{***}&     0.00939\sym{***}&     -0.0491\sym{***}&     -0.0475\sym{***}&     0.00979\sym{***}&     0.00920\sym{***}&      0.0190\sym{***}&      0.0184\sym{***}\\
                    &  (0.000503)         &  (0.000503)         &   (0.00252)         &   (0.00252)         &   (0.00145)         &   (0.00145)         &  (0.000706)         &  (0.000705)         \\
[0.9em]
\# MeSH             &    -0.00476\sym{***}&    -0.00697\sym{***}&    -0.00823\sym{***}&     0.00671\sym{**} &    -0.00451\sym{***}&    -0.00857\sym{***}&     0.00420\sym{***}&    0.000859         \\
                    &  (0.000403)         &  (0.000431)         &   (0.00194)         &   (0.00207)         &   (0.00111)         &   (0.00119)         &  (0.000542)         &  (0.000577)         \\
\end{tabular}
\end{table*}
\end{landscape}
\restoregeometry

\end{document}